\def\tr{\mathop{\rm tr}\nolimits}

\documentstyle [preprint,aps,eqsecnum]{revtex}
\begin{document}
\begin{center}
{\bf A CONSTITUENT QUARK ANTI-QUARK EFFECTIVE LAGRANGIAN BASED ON THE DUAL
SUPERCONDUCTING MODEL OF LONG DISTANCE QCD}\\
\vspace{.1in}
M. Baker\footnote{Work supported in part by the U.S. Dept. of Energy under
Contract No. DE-FG06-91ER40614}\\
{\it University of Washington, Seattle, WA 98105}\\
James S. Ball\\
{\it University of Utah, Salt Lake City, UT 84112}\\
F. Zachariasen \footnote{Work supported in part by the U.S. Dept. of Energy
under Grant No. DE-FG03-92-ER40701}\\
{\it California Institute of Technology, Pasadena, CA 91125}\\
\end{center}
\vspace{.5in}
\begin{center}
{\bf ABSTRACT}
\end{center}

We review the assumptions leading to the description of long distance QCD by a
Lagrangian density expressed in
terms of dual potentials.  We find the color field distribution surrounding a
quark anti-quark pair to first order in their velocities.  Using these
distributions we eliminate the dual potentials from the Lagrangian density and
obtain an effective interaction Lagrangian $L_I ( \vec x_1 \, , \vec x_2 \, ;
\vec v_1 \, ,
\vec v_2 )$ depending only upon the quark
and anti-quark coordinates and velocities, valid to second order in their
velocities.  We propose $L_I$ as the Lagrangian describing the long distance
interaction between constituent quarks. Elsewhere we have determined the two
free parameters in $L_I$, $\alpha_s$ and the string tension $\sigma$, by
fitting the 17 known levels of $b \bar b$ and $c
\bar c$ systems.  Here we use $L_I$ at the classical level to calculate the
leading Regge trajectory.  We obtain a trajectory which becomes linear at
large $M^2$ with a slope $\alpha' \simeq .74 \, \hbox{GeV}^{-1}$, and for small
$M^2$ the trajectory
bends so that there are no tachyons.  For a constituent quark mass between 100
and 150 MeV this trajectory passes through the two known Regge recurrences of
the $\pi$ meson.  In this paper, for simplicity of presentation, we have
treated
the quarks as spin-zero particles.

\newpage

\section{\bf INTRODUCTION}

In 1964 Gell-Mann and Zweig proposed that the constituents of hadrons are
quarks u, d, and s which have masses of order of 300 MeV and which carry
color degrees of freedom.  Elementary calculations analogous to those used in
atomic physics give a good qualitative description of the hadronic spectrum.
We know that the forces in an atom are mediated by the electromagnetic field
$A_{\mu}$.  One might ask what is the field outside constituent quarks which
mediates their interaction?  From the point of view of QCD constituent quarks
are superpositions of interacting gluons $A^{a}_{\mu}$ and fundamental spin
$1/2$ quark fields $\psi_{\alpha}$.  At short distances where the coupling is
weak (asymptotic freedom) one can use the Yang Mills potentials $A^a_{\mu}$
and carry out perturbation calculations.  At long distances the coupling
becomes strong (anti-screening) and there is no procedure available to use
the potentials $A^a_{\mu}$ to calculate the force between quarks.

In this paper we propose that dual potentials $C^a_{\mu}$ mediate the long
distance interactions between constituent quarks.%
\footnote%
{We have previously discussed the motivation for using dual potentials to
describe long distance QCD.  In order to make this paper self-contained, we
present here the arguments and assumptions involved without reference to our
earlier work.}
We construct an effective
Lagrangian density ${\cal L}_{\hbox{\tiny eff}}$ to describe the $C^{a}_{\mu}$
quark
interactions, treating the quarks as classical point particles having
coordinates $\vec{x}_{i}(t)$ and velocities $\vec{v}_{i}(t)$.  We solve the
resulting classical field equations for $C^{a}_{\mu}$ in the presence of a
quark anti-quark pair to first order in their velocities.  Using these
solutions we eliminate the dual potentials $C^{a}_{\mu}$ from ${\cal
L}_{\hbox{\tiny eff}}$ obtaining an effective interaction Lagrangian $
L_I(\vec
{x}_{1},\vec{x}_{2}; \vec{v}_{1}, \vec{v}_{2}) \equiv \int d\vec{x}{\cal
L}_{\hbox{\tiny eff}}(\vec{x})$ depending upon the quark and anti-quark
positions and
velocities valid to second order in these velocities.  In another paper
\cite{1},
starting with $L_{I}$, we carry out the canonical quantization procedure to
construct the Hamiltonian for an interacting quark anti-quark pair and
calculate the spectrum of mesons composed of heavy quarks.  In this paper we
use the effective Lagrangian $L_{I}$ at the classical level to calculate Regge
trajectories.  For the sake of the simplicity of this presentation we do not
treat the spin degrees of freedom here. (These have been considered elsewhere
\cite{1} \cite{2}).  The
principal new development is the treatment of the quark's
coordinates and velocities as dynamical variables rather than as externally
prescribed parameters.  This makes possible the construction of the effective
Lagrangian $L
_{I}$ depending only upon particle coordinates and velocities.  The
same procedure can be used to calculate the effective Lagrangian $L_{I}
(\vec{x}_{1}, \vec{x}_{2}, \vec{x}_{3}; \vec{v}_{1}, \vec{v}_{2},
\vec{v}_{3})$ for the system of three interacting quarks in a baryon.

\section{\bf  ABELIAN THEORY}

We first review how abelian dual potentials $C_{\mu}$ provide an alternate
framework to solve Maxwell's equations in a linear relativistic medium having
a wave number dependent dielectric constant $\varepsilon (q)$ and magnetic
permeability $\mu (q)=1/{\varepsilon (q)}$ in the presence of a given charge
density $\rho (\vec{x})$ and current density $\vec{j}(\vec{x})$ \cite{3}.  We
first
express $\rho$ and $\vec{j}$ in terms of lines of polarization (Dirac strings)
attached to the moving charged particles, i.e. we find a string polarization
$\vec{P}_s$ and a string magnetization $\vec{M}_s$ so that

\begin{equation} \label{1}
\rho = - \vec{\nabla}\cdot \vec{P}_s\quad ,\quad \vec{j} = {\partial \vec{P}_s
\over\partial t} +
\vec{\nabla}\times\vec{M}_s\,.
\end{equation}

The solution to eq.~(\ref{1}) for the case of two particles of charges $e(-e)$
moving along trajectories $\vec{x}_1(t) (\vec{x}_2(t))$ is

\begin{equation} \label{2}
\vec{P}_s (\vec{x},t)= e\int_{\vec{x}_{2}(t)}^{\vec{x}_{1}(t)} d\vec{y}
\delta(\vec{x} - \vec{y}) \, ,
\end{equation}

\begin{equation} \label{3}
\vec{M}_{s}(\vec{x},t) = e \int_{\vec{x}_{2}(t)}^{\vec{x}_{1}(t)}
d\vec{y}\times\dot{\vec{y}}\delta (\vec{x}-\vec{y}) \, ,
\end{equation}
\noindent
where the integrals $\vec{d}y$ are along any path $L(t)$ connecting
$\vec{x}_{1}(t)$
and $\vec{x}_{2}(t)$ and $\dot{\vec{y}}$ is the velocity of the portion of the
path between $\vec{y}$ and $\vec{y} + d \vec y$. (See Fig. 1).
It is readily checked that the string polarization $\vec{P}_{s}$ and
magnetization $\vec{M}_{s}$ yield eq.~(\ref{1}) with

\begin{equation} \label{4}
\rho = e [\delta (\vec{x} - \vec{x}_{1}(t)) - \delta ( \vec x - \vec x_2 (t))]
\, ,
\end{equation}

\begin{equation} \label{5}
\vec{j} = e [\vec{v}_{1}(t) \delta (\vec{x} -\vec{x}_{1}(t)) - \vec{v}_{2}(t)
\delta (\vec{x} - \vec{x}_{2}(t))] \, .
\end{equation}
For example taking the divergence of eq.~(\ref{2}) yields
\begin{equation} \label{6}
\vec{\nabla}\cdot\vec{P}_{s} = e \int_{\vec{x}_{2}(t)}^{\vec{x}_{1}(t)}
d\vec{y}\cdot \vec{\nabla}\delta (\vec{x} - \vec{y}) = -\rho\, .
\end{equation}
Inserting expressions (\ref{1}) for $\rho$ and $\vec{j}$ into Gauss' Law,
$\vec{\nabla}\cdot\vec{D} = \rho$, and Ampere's Law, $\vec{\nabla}\times\vec{H}
=
\vec{j} + {\partial\vec D\over\partial t}$, yields equations for $\vec{D}$
and $\vec{H}$ whose general solution is
\begin{equation} \label{7}
\vec{D} = -\vec{\nabla}\times\vec{C} - \vec{P}_{s} \,,  \qquad \qquad
\vec{H} = -
{\partial\vec{C}\over\partial t} - \vec{\nabla}C_{0} + \vec{M}_{s} \, .
\end{equation}
Thus Gauss' Law and Ampere's Law have become kinematical equations whose
solution defines dual potentials $C^{\mu} \equiv (C_0 \, , \vec C )$.

The homogeneous Maxwell equations, $\vec{\nabla}\cdot\vec{B} =
0$\, ,$\vec{\nabla}\times\vec{E} = -\frac{\partial\vec{B}}{\partial t}$, taken
together with the constitutive equations, $ \vec{B} = \mu(q) \vec{H} $ , $
\vec{E} =
\frac{1}{\epsilon (q)} \vec{D} = \mu (q)\vec D$ and eq.~(\ref{7}) for $\vec{D}$
 and
$\vec{H}$ then yield the following dynamical equations for the dual potentials
$C_{\mu}$:
\begin{equation} \label{8}
\partial^\alpha \mu (\partial_\alpha C_\beta - \partial_\beta C_\alpha ) =
- \partial^{\alpha}(\mu G^{s}_{\alpha\beta}) \, ,
\end{equation}
where the string polarization tensor $G_{\alpha\beta}^{s}$ appearing in the
inhomogeneous term in eq.~(\ref{8}) has the following components:
\begin{equation} \label{9}
G_{0k}^{s} = M_{s,k}\, ,\hspace{.5in}G_{ij}^s  = -\epsilon_{ijk}P_{s,k} \, .
\end{equation}
Eqs. (\ref{8}) provide an alternate form of Maxwell's equations, which are
completely equivalent to their usual form expressed in terms of ordinary
potentials $A_{\alpha}$,
\begin{equation} \label{10}
\partial^{\alpha}\epsilon
(\partial_{\alpha}A_{\beta}-\partial_{\beta}A_{\alpha}) = J_{\beta},
\end{equation}
where $J^\beta \equiv (\rho ,\vec j)$\, .
We now show how eqs.~(\ref{2}), (\ref{3}), (\ref{7}), (\ref{8}) and (\ref{9})
with $\mu = 1$ give the usual
electric and magnetic fields of slowly moving particles.

First consider the
case of particles at rest.  Then the space component of eq.~(\ref{8}),
$(\vec{\nabla}\times\vec{E} = 0)$, determining the dual vector potential
(denoted $\vec{C}_{\hbox{\tiny DIRAC}} \equiv \vec{C}_{D}$) takes the form
\begin{equation} \label{11}
\vec{\nabla}\times (-\vec{\nabla}\times \vec{C}_{D}) =
\vec{\nabla}\times\vec{P}_{s}\, .
\end{equation}
Comparing eq.~(\ref{11}) with the equation determining the vector potential of
a
magnetic dipole and using eq.~(\ref{2}), we obtain the solution of eq.~
(\ref{11}):
\begin{equation} \label{12}
\vec C_D = -\int_{\vec{x}_{2}(t)}^{\vec{x}_{1}(t)}
{ed\vec{y}\over 4 \pi}\times{(\vec x - \vec y) \over {\mid\vec x - \vec y
\mid}^3} \, .
\end{equation}
Eqs. (\ref{2}), (\ref{7}) and (\ref{12}) then yield the electric field
\begin{equation} \label{13}
\vec{D} = -\vec{\nabla}\times\vec{C}_{D} - \vec{P}_{s} =
\vec{D}_{\hbox{\tiny COULOMB}} \equiv \vec{D}_{\hbox{\tiny C}} \, ,
\end{equation}
\noindent
where
\begin{equation} \label{14}
\vec{D}_{\hbox{\tiny C}} = \frac{e}{4\pi}\left \{\frac{\vec x -\vec
x_1}{{\mid\vec x - \vec x_1 \mid}^3} -
{\vec x - \vec x_2\over{\mid\vec x - \vec x_2 \mid}^3}\right \} \, .
\end{equation}
The first term $-\vec{\nabla}\times\vec{C}_{D}$ in eq.~(\ref{13}) produces a
field
analogous to the magnetic field of a line of magnetization, which is the sum
of a Coulomb field and field flowing through the string to provide the
continuity of the field lines.  The second term $\vec{P}_{s}$ cancels
the field through the string leaving the desired pure Coulomb field.  See Fig.
2.

For slowly moving charges the time component of eq.~(\ref{8})
$(\vec{\nabla}\cdot\vec{B} = 0)$, determining the scalar potential (denoted
$C_{0D})$ takes the form:
\begin{equation} \label{15}
-\nabla^{2}C_{0D} = -\vec{\nabla}\cdot\vec{M}_{s}\, .
\end{equation}
Comparing eq.~(\ref{15}) with the equation for the scalar potential of an
electric dipole, and using eq.~(\ref{3}), we obtain the solution of eq.~
(\ref{15}):
\begin{equation} \label{16}
C_{0D} = \int_{\vec{x}_{2}(t)}^{\vec{x}_{1}(t)}
{ed\vec{y}\over 4 \pi}\times\dot{\vec{y}}\cdot{(\vec x - \vec y) \over
{\mid\vec{x}-\vec{y}\mid}^3}
\, .
\end{equation}
Evaluating the magnetic field $\vec{H}$ from eq.~(\ref{7}), and using
eqs.~(\ref{3}), (\ref{12}) and
(\ref{16}), we obtain
\begin{equation} \label{17}
\vec H = - \vec{\nabla} C_{0D} - \frac{\partial\vec C_D}{\partial t} +
\vec M_s = \vec H_{\hbox{\tiny BIOT SAVART}} \equiv \vec H_{BS}\,,
\end{equation}
where
\begin{equation} \label{18}
\vec H_{BS} = \frac{e}{4\pi} \left[ {\vec v_1 \times (\vec x - \vec
x_1) \over {| \vec x - \vec x_1 |}^3} - {\vec v_2 \times (\vec x - \vec
x_2) \over {| \vec x - \vec x_2 |}^3} \right]
\end{equation}
is the usual Biot Savart magnetic field produced by slowly moving charges.

Eq. (\ref{1}) can be written in the covariant form
\begin{equation} \label{19}
J_\alpha = - \frac{1}{2} \partial^\beta
\epsilon_{\alpha\beta\sigma\lambda}G^{s,\,\sigma\lambda}\,,
\end{equation}
which makes manifest the electromagnetic duality connecting eqs.~(\ref{8}) and
(\ref{10}):
\begin{equation} \label{20}
C_\mu \longleftrightarrow A_\mu \,, \quad  \mu = {1 \over \epsilon}
\longleftrightarrow \epsilon \,, \quad
\mu G_{\lambda \mu}^s \longleftrightarrow \frac{1}{2}
\epsilon_{\lambda \mu \alpha \beta} G^{s,\, \alpha \beta}\,.
\end{equation}
All textbooks on electricity and magnetism could be rewritten using only dual
potentials $C_\mu $ satisfying eq.~(\ref{8}) and the same electromagnetic
forces
between charged particles would be obtained.  The potentials themselves
however could be completely different from the conventional $A_{\mu}$.  For
example in a dielectric medium
where $\epsilon (q) \rightarrow 0$ as $q \rightarrow 0$, corresponding to
anti-screening at large distances, the potential $A_\mu$ determined from
eq.~(\ref{10}) would be singular at large distances while the dual potentials
$C_\mu$
satisfying eq.~(\ref{8}) with $\mu = \frac{1}{\epsilon} \rightarrow \infty$ as
$q^2
\rightarrow 0$ would be screened at large distance.  Use of the $A_\mu$ to
describe this system would introduce singularities which do not appear in the
dual potentials $C_\mu$.  These are thus the natural choice of potentials
to describe a medium with long range anti-screening.

\section{\bf  NON ABELIAN THEORY (WITHOUT QUARKS)}

The main point of the above discussion is that there is an additional degree
of freedom available for the description of gauge theories beyond the usual
freedom to make gauge transformations on the potentials $A_\mu$.  We will
exploit this freedom in our study of non abelian gauge theories.  In
Yang Mills theory the effective dielectric constant results from the non
linear interaction of the vector potentials $A_\mu$ leading to asymptotic
freedom, i.e. anti-screening.  We want to find an effective Lagrangian
${\cal L}_{\hbox{\tiny eff}}$
to describe the long distance behavior of such a medium in terms of non abelian
dual potentials $C^a_\mu$.  Mandelstam\cite{3A} has given a formal relation
between
the non abelian dual potentials $C^a_\mu$ and the Yang Mills potentials
$A^a_\mu$
and 'tHooft \cite{3B} has shown that if Yang Mills theory confines quarks then
the dual Wilson loop
(that defined in terms of dual potentials) satisfies a perimeter law.
(This is the gauge invariant non abelian analogue of the correspondence,
eq.~(\ref{20}), obtained in abelian theory showing that in a medium with
anti-screening
the dual potentials are non singular at large distances.)  To guarantee this
smooth long distance behavior of the dual potential we then require \bigskip

\noindent  {\bf a)} ${\cal L}_{\hbox{\tiny eff}}$ must generate a mass for the
$C^a_\mu$ field.  This mass provides the
shielding of the dual potentials expected from the long distance extrapolation
of asymptotic freedom.  \bigskip

\noindent  {\bf b)} ${\cal L}_{\hbox{\tiny eff}}$ must be invariant under non
abelian gauge transformations of the
dual potentials. (${\cal L}_{\hbox{\tiny eff}}$ must incorporate the
fundamental color symmetry of
QCD.)
We write this gauge transformation in the usual matrix form,
\begin{equation} \label{21}
C_\mu = \Omega^{-1}C_\mu \Omega + {i \over g} \Omega^{-1}\partial_\mu \Omega\,
,
\end{equation}
where $C_\mu = \sum_{a = 1}^{8} C^a_\mu {1 \over 2}\lambda_a $ \,, $ \Omega$ is
an
$SU(3)$ matrix and $g = {2\pi \over e}$ where e is the Yang Mills coupling
constant, i.e. $\alpha_s = {e^2 \over 4\pi}$. (For $SU(N)$ gauge theory $C_\mu$
and
$\Omega$ are $SU(N)$ matrices, and ${1 \over 2}\lambda_a$ are the generators
$SU(N)$, normalized so that $2 \tr$ $\lambda_a \lambda_b = \delta_{ab}$).
The relation $ g = {2\pi \over
e} $ between the coupling $g$ of the dual potential $C^a_\mu$ and the Yang
Mills
coupling constant $e$ can be obtained from 'tHooft's duality relations
\cite{3B} between
non abelian electric and magnetic flux.
We will write down a minimal Lagrangian ${\cal L}_{\hbox{\tiny eff}}$
satisfying (a) and (b).
There are in general an infinite number of terms which can be added to this
minimal ${\cal L}_{\hbox{\tiny eff}}$ compatible with the requirements (a) and
(b).  However
contributions to ${\cal L}_{\hbox{\tiny eff}}$ involving higher dimension
operators should be
surpressed at long distances and we retain only the lowest dimension
operators.  Our third assumption is then \bigskip

\noindent  {\bf c)}  Higher dimension operators not included in ${\cal
L}_{\hbox{\tiny eff}}$ are not quantitatively
important for describing long distance phenomena.

To construct ${\cal L}_{\hbox{\tiny eff}}$ we use the Higgs mechanism to
generate the $C^a_\mu$
mass and couple $C^a_\mu$ to three scalar Higgs fields $B^a_1  , B^a_2  ,
B^a_3$,
$ a = 1,2,\cdot\cdot\cdot N^2 -1$, in the adjoint representation
of $SU(N)$.  The additional effective fields
$B^a_i\,, i = 1,2,3\,,$ are needed in order to render ${\cal L}_{\hbox{\tiny
eff}}$ local.  Since they couple in a
minimal way to the dual potentials they must carry magnetic charge.  There can
not be any fields in ${\cal L}_{\hbox{\tiny eff}}$ which transform like the
fundamental
representation of $SU(N)$ since ${\cal L}_{\hbox{\tiny eff}}$ must have the
same symmetry as Yang
Mills theory.  (We introduce the coupling to quarks later.)  We include in
${\cal L}_{\hbox{\tiny eff}}$ a Higgs potential $W(B)$ which has a minimum at a
non vanishing
value of $B^a_i$.  For $SU(3)$ , $B^a_i$ are octets
defining three $SU(3)$ matrices $B_i = \sum^8_{a = i} B^a_i {\lambda_a \over 2}
$,
 and the vacuum expectation values $B_{0i}$ of $B_i$ have the
following structure:
\begin{equation} \label{22}
B_{01} = B_0 \lambda_7 \,\,,  B_{02} = B_0 (-\lambda_5)\, \,, B_{03} = B_0
\lambda_2 \,,
\end{equation}
where the value of $B_0$ is determined by the position of the minimum in the
Higgs potential $W$.  The three matrices $ \lambda_7$, $ - \lambda_5 $, and $
\lambda_2 $ transform as a $ j = 1 $ irreducible representation of an $ SU(2)
$ subgroup of $ SU(3) $ and it is readily shown that there is no $ SU(3) $
transformation which leaves all three $ B_{0i} $ invariant.  The $ SU(3) $
gauge symmetry is then completely broken and the eight Goldstone bosons become
the longitudinal components of the now massive $ C_\mu $. (The ansatz
(\ref{22})
is readily extended to $ SU(N) $ such that the $SU(N)$ gauge symmetry
is completely broken.  However in the rest of this paper we restrict ourselves
to the case of interest, namely $SU(3)$).

The Lagrangian $ {\cal L}_{\hbox{\tiny eff}} $ takes on the
following form:

\begin{equation} \label{23}
   {\cal L}_{\hbox{\tiny eff}} = 2 \tr \left\{ - \frac{1}{4}  G^{\mu\nu}
G_{\mu\nu} +\frac{1}{2}
  {({\cal D}_\mu B_i)}^2 \right\} - W(B) + \cdot \cdot \cdot \, ,
\end{equation}
where
\begin{equation} \label{24}
   {\cal D}_\mu B_i = \partial_\mu B_i - ig [ C_\mu , B_i ] \, ,
\end{equation}
\begin{equation} \label{25}
   G_{\mu\nu} = \partial_\mu C_\nu - \partial_\nu C_\mu - ig [ C_\mu , C_\nu ]
   \, .
\end{equation}
The explicit color structure of $ W $ is \cite{3C}
\begin{equation} \label{26}
   W = 3 \mu^2 \tr B_i^2 + {2 \over 3} \lambda \{ 3 \tr B_i^2 B_j^2 + (\tr
   B_i^2)^2 + 2 (\tr B_i B_j)^2 \} \, ,
\end{equation}
with $ \lambda > 0 $ and $ \mu^2 < 0 $ so that $ W$ has a minimum at a non
vanishing value of $ B_0 = 3 \mu /8 \sqrt{\lambda} $, which determines the
vacuum $B_i$ fields $B_{0i}$ via eq.~(\ref{22})\footnote{The expression
(\ref{26}) is the minimal counterterm
necessary for the renormalizability of  $ {\cal L}_{\hbox{\tiny eff}} $
\cite{3C}.  For simplicity we have not included other operators of the same
dimension.}.
  As indicated above there are an infinite number of
terms containing higher dimensional operators which can be added to
eq.~(\ref{23}),
but are not included by assumption (c).  The theory then contains three
parameters: the dual gauge coupling constant $ g = {2 \pi \over e } $, the
value $ B_0 $ of the vacuum expectation value of the Higgs field and $
\lambda $, the quartic Higgs coupling. The non abelian Lagrangian $ {\cal
L}_{\hbox{\tiny eff}} $,
eq.~(\ref{23}), resembles the Lagrangian of the abelian Higgs model.  We will
see
that it describes a dual superconductor and provides a concrete realization of
the Mandelstam-'tHooft dual superconductor mechanism of confinement \cite{4}
\cite{5}.

The basic manifestation of the dual superconducting properties of $ {\cal
L}_{\hbox{\tiny eff}} $
is that it generates classical equations of motion having static electric flux
tube solutions carrying a unit of $ Z_3 $ flux along the $z$ axis \cite{6}.
These are
the non abelian analogues of the Nielsen-Olesen magnetic vortex solutions of
the
abelian Higgs model \cite{6A}. We choose the gauge so that $ C_0 = 0 $, $
\vec{\nabla} \cdot
\vec C = 0 $, and $ \vec C $ is proportional to  \^{e}$_\phi $, the unit vector
along the $ \phi $ direction and to the hypercharge matrix $ Y = {\lambda_8
\over \sqrt{3} } $.  The Higgs fields $ B_i $ lie along the same directions in
color space as their vacuum values $ B_{0i} $, eq.~(\ref{22}), i.e.
\begin{equation} \label{27}
   B_1 (\vec x) = \lambda_7 B (\vec x)\, , \quad  B_2 (\vec x)  = - \lambda_5
B (\vec x)
  \, , \quad B_3 (\vec x) = \lambda_2 B^{'} (\vec x )\, .
\end{equation}
At large distances $\rho $ from the flux tube $ B_i (\vec x ) \rightarrow
B_{0i} $.  The non vanishing of $ B_{0i} $ produces a color
monopole current confining the electric flux.  Since $ \vec C $ is
proportional to $ Y $, the non abelian term in eq.~(\ref{25}) vanishes.  We
then have
$ G_{0i} = 0 $, $ G_{ij} = \epsilon_{ijk} D_k $, where the color electric
field,
$ \vec D = - \vec{\nabla} \times \vec C $, is proportional to the hypercharge $
Y $.  The color electric flux $ \Phi_S $ passing through a large circle $S$
perpendicular to the $ z $ axis is then
\begin{equation} \label{28}
   \Phi_S = \int_S d \vec a \cdot \vec D = - \oint d \vec l \cdot \vec C \,,
\end{equation}
where the line integral in eq.~(\ref{28}) runs over the perimeter of $S$.
As $ \rho \rightarrow \infty $, $ \vec C \rightarrow - { \hat e_\phi Y \over g
\rho } $, and eq.~(\ref{28}) gives
\begin{equation} \label{29}
 \Phi_S = {2\pi \over g} Y = e Y \, ,
\end{equation}
or
\begin{equation} \label{30}
    e^{ig\Phi_S} = e^{2 \pi i Y} = e^{{(2 \pi i)
   \over 3}} \, .
\end{equation}
A continuous deformation in $ SU(3) $ of our particular solution into a non
abelian configuration will leave unchanged $ P \hbox{exp} (ig \int \vec C \cdot
d
\vec l) = e^{{2\pi i\over 3}} $. The flux tube then contains one unit of $
Z_3 $ flux.  The energy per unit length in this flux tube defines the string
tension $\sigma$. The quantity $g^2/\lambda$ in ${\cal L}_{\hbox {\tiny eff}}$
plays the role of a Landau Ginzburg parameter.  It was determined by using the
trace anomaly relating the difference of the value of energy density at large
distances from the flux tube and its value at the center of the flux tube to
the gluon condensate $G_2$.  The result \cite{8}  $g^2 /\lambda = 5 $ is near
the
border between type I and type
II superconductors.  We
obtain from the numerical integration of the static field equations with $g^2
/\lambda = 5$:
\begin{equation} \label{A1}
    \sigma = (4.9 B_0 )^2 \, .
\end{equation}
Having fixed the ratio $g^2/\lambda $ of the parameters appearing in ${\cal
L}_{\hbox{\tiny eff}}$ we are left with two free parameters to determine the
quark anti-quark Lagrangian $L_I$.  These can be taken as $\sigma$ and
$\alpha_s \equiv {e^2 \over 4 \pi} = {\pi \over g^2}$.

\section{\bf  THE EFFECTIVE QUARK ANTI-QUARK LAGRANGIAN}

To couple $ C_\mu $ to a quark anti-quark pair we introduce a Dirac string
connecting the pair and represent their charge density $ \rho $ and current
density $\vec{j}$ as polarization charge and polarization current as in
eqs.~(\ref{1}), (\ref{2})
and (\ref{3}).  The only difference is that $ \rho $ and $ \vec j $ are
proportional
to the hypercharge matrix $ Y $ so that the quark (anti-quark) can emit
(absorb) the unit of $ Z_3 $ flux described by the flux tube solution.  The
polarization $ \vec P_s $ and magnetization $ \vec M_s $ and hence
$ C_{\mu} $ are proportional to $Y$. The non abelian contribution to
$G_{\mu\nu}$, (\ref{25}) vanishes and the coupling of dual potentials to quarks
proceeds as in the abelian case.  All quantities $ C_\mu \, , \vec
D \, , \vec H \, , \vec j \, , \rho $ appearing in eqs.~(\ref{1}) -(\ref{7})
are
automatically proportional to $ Y $, and these equations are otherwise
unchanged.  We now redefine $G_{\mu\nu}$:
\begin{equation}\label{1new}
   G_{0k} \equiv H_k\, , \quad G_{ij} \equiv \epsilon_{ijk} D_k \, ,
\end{equation}
and eq.~(\ref{7}) becomes
\begin{equation} \label{31}
   G_{\mu\nu} = ( \partial_\mu C_\nu - \partial_\nu C_\mu + G_{\mu\nu}^s )\,
   ,
\end{equation}
where $G^s_{\mu\nu}$ is given by eq.~(\ref{9}).  We then obtain the Lagrangian
density ${\cal L}_{\hbox{\tiny eff}}$ describing the coupling of dual gluons
to a quark anti-quark pair by inserting eq.~(\ref{31}) for $G_{\mu\nu}$ into
eq.~(\ref{23}).  This yields:
\begin{equation} \label{32}
   {\cal L}_{\hbox{\tiny eff}} = 2 \tr \left \{- {1 \over 4}(\partial_\mu
C_\nu - \partial_\nu C_\mu + G_{\mu\nu}^s )^2
    + {1 \over 2}({\cal D}_\mu B_i)^2  \right \} - W \, .
\end{equation}

We make the same color ansatz, eq.~(\ref{27}), for $B_i$.
Varying $ C_\mu $ in $ {\cal L}_{\hbox{\tiny eff}} $ gives the equations of
motion:
\begin{equation} \label{33}
   \partial^\mu ( \partial_\mu C_\nu - \partial_\nu C_\mu ) =
  - \partial^\mu G_{\mu\nu}^s + J^B_\nu \, ,
\end{equation}
where
\begin{equation} \label{34}
   J^B_\nu \equiv i g [ {\cal D}_\nu B_i\, , B_i ] = 6 g^2 C_\nu B^2 (x) \, .
\end{equation}
We have used the color structure (\ref{27}) of $B_i$ to obtain the second
equality in eq.~(\ref{34}). All terms in eq.~(\ref{33}) are in the $ Y $ color
direction.
The first term on the right
hand side of eq.~(\ref{33}) is the polarization current $ \partial^\mu
G_{\mu\nu}^s $ of the moving Dirac string.  Taken by itself it would
produce color fields satisfying Maxwell's equations (eq.~(\ref{8}) with $ \mu =
1
$). The second term $ J_\nu^B $ is the monopole current carried by the
fields $ B_i $ and arises from the non abelian terms in eq.~(\ref{32}).
It shields the dual potentials so that the color flux emanating from the pair
is confined.  (In a sense the non abelian coupling of $
C_\mu $ to $ B_i $ in eq.~(\ref{33}) plays the role of the magnetic
permeability$\mu$ in eq.~(\ref{8})).  Varying $B_i$ gives the equations for
$B_i$, namely
\begin{equation} \label{35}
   {\cal D}_\mu {\cal D}^\mu B_i = {1 \over 2} { \delta W \over \delta B_i }
   \,,
\end{equation}
which have components in the $\lambda_2$, $\lambda_5$ and $\lambda_7$
directions in color space.

The solutions of equations (\ref{33}) and
(\ref{35}) depend parametrically upon $ \vec R (t) = \vec x_1 (t) - \vec x_2
(t) $, $ \vec v_1 $ and $ \vec v_2 $.
We will see that to first order in $\vec v_1$ and $\vec v_2$ the time
dependence of $C_\mu$ and $B_i$ appears only via the explicit time dependence
of $\vec R$, $\vec v_1$ and $\vec v_2$, i.e.,
\begin{equation} \label{36}
   C_\mu (\vec x,t ) = C_\mu ( \vec x \, , \vec R \, , \vec v_1 \, , \vec v_2 )
     ,\quad
	 B_i (\vec x,t ) = B_i ( \vec x \, , \vec R \, , \vec v_1 \, , \vec v_2
	 )\,.
 \end{equation}
Substituting the first order solutions (\ref{36}) into
${\cal L}_{\hbox{\tiny eff}}$ and thereby eliminating the field
variables $C_\mu$ and $B_i$ gives the effective quark anti-quark Lagrangian
\begin{equation} \label{37}
  L_I (\vec x_1 , \vec x_2 ; \vec v_1 , \vec v_2 ) = \int d \vec x  {\cal
     L}_{\hbox{\tiny eff}}  (C_\mu , B_i )\, ,
\end{equation}
    which we will see is valid to second order in the velocities.
      The effective interaction
Lagrangian $  L_I $ (\ref{37}) supplemented by the contribution of the quark
kinetic energy is our starting point for the  constituent quark model.  It
contains the
quark masses, $ B_0 $, $ \lambda $, and $ g = {2\pi \over e} $ as parameters.
However, $\lambda = g^2 /5 = {\pi \over 5 \alpha_s}$ and $ B_0 =
\sqrt{\sigma} /4.9$ (eq.~(\ref{A1})).  Thus the parameters appearing in $L_I$
are $\alpha_s$ and the string tension $\sigma$.

We now outline the procedure used to calculate $ L_I $.  First we express
$ {\cal L}_{\hbox{\tiny eff}} $ in terms of the color electric and magnetic
fields $ \vec D $ and $
\vec H $.  Using the fact that $ - ( \partial_\mu C_\nu - \partial_\nu C_\mu +
G_{\mu\nu}^s )^2 = 2 ( \vec H^2 - \vec D )^2 $, we can write eq.~(\ref{32}) as
\begin{equation} \label{38}
   {\cal L}_{\hbox{\tiny eff}} = {\cal L}_0 + {\cal L}_2 \, ,
\end{equation}
where
\begin{equation} \label{39}
   {\cal L}_0 = -2 \tr \left\{ \begin{array}{cc}{ \vec D^2 \over 2} +
   { ( \vec {\cal D} B_i )^2 \over 2}\end{array}  \right\} - W
   \, ,
\end{equation}
and
\begin{equation} \label{40}
   {\cal L}_2 = 2 \tr \left\{ \begin{array}{cc}{ \vec H^2 \over 2} +
   { ( {\cal D}_0 B )^2 \over 2 }\end{array} \right\} \, .
\end{equation}
All terms in  ${\cal L}_{\hbox{\tiny eff}} $ with time derivatives are
contained in $ {\cal L}_2 $ and for a
static configuration with $ C_0 = 0 $, ${\cal L}_{\hbox{\tiny eff}} = {\cal
L}_0 $.  The Lagrangian
$  L_I $ can then be written as
\begin{equation} \label{41}
   L_I = \int d \vec x {\cal L}_0 +\int d \vec x {\cal L}_2 \,.
\end{equation}
Next using eqs.~(\ref{9}) and (\ref{34}) we  write eq.~(\ref{33}) in the three
dimensional
form:
\begin{equation} \label{42}
   \vec{\nabla} \times ( \vec{\nabla} \times \vec C ) + \partial_0^2 \vec C +
   { \partial \over \partial t} \vec{\nabla} C_0 = - \vec{\nabla} \times
   \vec P_s + { \partial \vec M_s \over \partial t } - 6 g^2 B^2 \vec C \, ,
\end{equation}
\begin{equation} \label{43}
   - \nabla^2 C_0 - { \partial \over \partial t} \vec{\nabla} \cdot \vec C =
   - \vec{\nabla} \cdot \vec M_s - 6 g^2 B^2 C_0\,.
\end{equation}

For quarks at rest $ \vec v_1 = \vec v_2 = 0 $, $ C_0 = 0 $ and eqs.~(\ref{35})
and
(\ref{42}) become coupled static non linear equations for $ \vec C $ and $
B_i$.
Eq. (\ref{42}) becomes
\begin{equation} \label{44}
   \vec{\nabla} \times ( \vec{\nabla} \times \vec C ) + 6 g^2 B^2 \vec C =
   - \vec{\nabla} \times \vec P_s \, ,
\end{equation}
which differs from eq.~(\ref{11})
by the presence of the monopole
current $ 6g^2 B^2 \vec C $.  We make the
dependence of the solutions to these static equations on the particle
positions explicit, writing
\begin{equation} \label{45}
   \vec C = \vec C ( \vec x - \vec x_1 \,\, , \vec x - \vec x_2 ) \, , \quad
   B_i = B_i ( \vec x - \vec x_1 \, , \vec x - \vec x_2 ) \, .
\end{equation}
For $ g = 0 $, eq.~(\ref{44}) reduces to eq.~(\ref{11}) and $ \vec C = \vec C_D
$, eq.~(\ref{12}),
multiplied by $Y$.
For $ g \neq 0 $ we solve eqs.~(\ref{35}) and (\ref{44}) numerically \cite{7}.
At
large distances the monopole current screens
$ \vec C $ and
 $ \vec C$ has the behavior:
\begin{eqnarray} \label{46}
   & \vec C ( \vec x) & \longrightarrow  \vec C_D ( \vec x ) Y\, ,\quad
\hbox{as}
   \,
     \vec x \rightarrow 0  \nonumber \, ,\\
      & & \longrightarrow  \hbox {decreases exponentially as} \,
       \vec x \rightarrow \infty \, .
\end{eqnarray}
while
\begin{equation} \label{47}
   B_i (\vec x) \rightarrow B_{0i} \, , \quad \hbox{as} \, \vec x \rightarrow
   \infty \, .
\end{equation}

We insert these static solutions into ${\cal L}_0$ to obtain the static
potential $V_0 (R)$:
\begin{equation} \label{48}
   V_0 (R) = - \int d \vec x {\cal L}_0 ( \vec C\, , B_i )\, .
\end{equation}
For $ g =0$, $\vec D = \vec D_{\hbox{\tiny C}} Y$ and
\begin{equation} \label{49}
 V_0 (R)  =  2 \tr Y^2
  \int d \vec x { (\vec D_{\hbox{\tiny C}})^2 \over 2 }
    =  - {4 \over 3} {\alpha_s \over R } \, .
\end{equation}
The quark self energies as well as the vacuum energy $ W(B_0)$ have been
subtracted out in eq.~(\ref{49}).  For $ g \neq 0$ $V_0 (R)$ has the behavior
\begin{eqnarray} \label{50}
    V_0 (R) & \longrightarrow & -{4 \over 3} {\alpha_s \over R}\, , \quad
\hbox{as} \, R \rightarrow
    0 \, ,  \nonumber \\
     & & \nonumber\\
    & \longrightarrow & \sigma R \, , \quad \hbox{as} \, R \rightarrow \infty
\, ,
\end{eqnarray}
where $\sigma$ is the string tension determined by the flux tube solution.
The color electric field
$ \vec D $ evolves from a dipole field as $ R \rightarrow 0 $ to a
flux tube distribution for $ R \rightarrow \infty $.  $ V_0 (R) $ is
approximately the sum of a screened Coulomb potential and a linear  potential
since the dual potential is screened at distances $ R $ greater than the
screening length $ \left( \begin{array}{c} { 1 \over 6 g^2 B^2_0 } \end{array}
\right)^{1/2} $. (See eq.~(\ref{44})).  For completeness we present an
analytic parametrization of $V_0(R)$, eq.~(\ref{48}), evaluated with the
numerical solutions (\ref{45}) of the static field eqs.~(\ref{44}) and
(\ref{35}) \cite{1}:
\begin{equation}\label{A2}
   V_0(R) = - {4 \over 3}{\alpha_s \over R} e^{- .511 \sqrt{{\sigma \over
   \alpha_s} }R} + \sigma R - 0.646 (\alpha_s \sigma)^{1/2} \, .
\end{equation}

Next consider quarks in motion.  To first order in the velocities $\vec v_1 \,
,
\vec v_2 $ of the quarks the static field configurations (\ref{45}) follow
adiabatically the positions $ \vec x (t) ( \vec x_2 (t)) $ of the quark
(anti-quark).  This follows from the fact that all terms involving time
derivatives
in eq.~(\ref{42}) for $\vec C $ and  eq.~(\ref{35}) for $ B_i $ are second
order in $ \vec
v_1 $ and $ \vec v_2 $.  Thus
replacing $ \vec x_1 $ by $ \vec x_1 (t) $ and $ \vec x_2 $ by $ \vec x_2 (t)
$ in the static solution eq.~(\ref{45}), we obtain the solution of these
equations to first order in the velocities:
\begin{equation} \label{51}
   \vec C  =  \vec C ( \vec x - \vec x_1 (t) \; , \vec x - \vec x_2
   (t)) \, , \quad B_i  =  B_i ( \vec x - \vec x_1 (t) \; , \vec x - \vec x_2
(t)) \,
   .
\end{equation}
Since the quantity $\int d \vec x {\cal L}_0$ generates the static field
equations, it is stationary about solutions to these equations.  Thus
evaluating it using eqs.~(\ref{51}) for $\vec C$ and $B_i$ gives an expression
valid to second order in the velocities and eq.~(\ref{48}) remains unchanged
to this order.  All the velocity dependence of $L_I$ then appears in ${\cal
L}_2$.

${\cal L}_2$ depends quadratically upon $\partial_0 \vec C$,
$\partial_0 B_i$, and $C_0$, all of which are first order in $\vec v_1$ and
$\vec v_2$.  To find $C_0$ to this order we solve eq.~(\ref{43}), with $\vec
C$ and $B$ replaced by their static values.  Eq. (\ref{43}) then becomes the
following linear equation for $ C_0$:
\begin{equation} \label{52}
   - \nabla^2 C_0 + 6 g^2 B^2 C_0 = - \vec{\nabla} \cdot \vec M_s \, .
\end{equation}
Eq. (\ref{52}) differs from eq.~(\ref{15}) by the presence of the monopole
charge $ 6 g^2 C_0 B^2$.  For $g = 0$ eq.~(\ref{52}) reduces to eq.~(\ref{15})
and
$C_0 = C_{0D}Y$. (See eq.~(\ref{12})).
For $g \neq 0$ we solve eq.~(\ref{52}) numerically and find the solution $C_0$,
which has the following behavior:
\begin{eqnarray} \label{53}
   C_0 ( \vec x ) & \longrightarrow  & C_{0D} (\vec x)Y\,\, \quad \hbox{as} \,
\vec x \rightarrow
   0 \, ,\\
   & \longrightarrow & \hbox{decreases exponentially as} \, \vec x \rightarrow
   \infty \nonumber \, .
\end{eqnarray}
The time derivatives $\partial_0  B$ and $\partial_0 \vec C$ are obtained
from the static solutions, attached to the moving particles. (See
eqs.~(\ref{51})).  For example,
\begin{equation} \label{54}
   \partial_0 B_i = ( \vec v_1 \cdot \vec{\nabla}_1 + \vec v_2 \cdot
   \vec{\nabla}_2) B_i (\vec x - \vec x_1 (t) \, , \vec x - \vec x_2 (t))\, .
\end{equation}

Insertion of these expressions for $C_0$, $\partial_0 B_i$, and $\partial_0
\vec
C$ in ${\cal L}_2$ gives the velocity dependent potential $V_2$:
\begin{equation} \label{55}
   V_2 = - \int d \vec x {\cal L}_2 ( \vec H, {\cal D}_0 B_i ) \, .
\end{equation}
For $g = 0$ , ${\cal D}_0 B_i = 0$ and $\vec H = \vec H_{BS} Y$ (see eq.~
(\ref{18})).  Then
\begin{equation} \label{56}
   V_2  = - 2 \tr Y^2 \int d \vec x {( \vec H_{\hbox{\tiny BS}} )^2
      \over
	 2 }
	      =   { 2 \over 3 } { \alpha_s \over R } \left \{ \vec v_1 \cdot \vec
	     v_2 +
		 { \vec v_1 \cdot \vec R \vec v_2 \cdot \vec R \over R^2 }
		 \right \}\, ,
\end{equation}
where the self energies have been subtracted out.
For $ g \neq 0$ eq.~(\ref{55}) gives the following expression for $V_2$
\begin{eqnarray} \label{57}
   V_2 & =  &{ [ \vec R \times {d \vec R \over d t }]^2  \over 4 R^2 } V_{+}
   (R)   + { [ \vec R \times ( \vec v_1 + \vec v_2 )]^2 \over 4 R^2} V_{-} (R)
   \\
      & & + { \vec R \cdot ( \vec v_1 + \vec v_2 ) ]^2 \over 4 R^2}
      V_{\parallel} (R) -
	 { [ \vec R \cdot { d \vec R \over d t }]^2 \over 4 R^2} ({ 4 \alpha_s
	   \over 3 R}) \, ,  \nonumber
\end{eqnarray}
where the ``potentials''  $ V_{+} (R) $ , $ V_{-} (R) $ , and $
V_{\parallel} (R) $
are moments of the field distributions $ \vec C $ , $ C_0 $ and $
\vec B_i $, determined by eq.~(\ref{55}).
The explicit formulae for the integrals defining these moments are
given in reference \cite{1}.  Using the
numerical solution of eq.~(\ref{52}) for $ C_0$ to evaluate the
integral for $V_+$ obtained from eq.~(\ref{55}), we find the analytical
parametrization \cite{1}:
\begin{equation}\label{A3}
   V_+ (R) = - { 2 \alpha_s \over 3 R} e^{- 1.137 \sqrt{ \sigma \over
   \alpha_s} R} - .208 \sigma R + 1.118 \sqrt{\alpha_s \sigma} \, .
\end{equation}
Furthermore in section V we show:
\begin{equation}\label{A4}
   V_{-} = - {1 \over 2} V_0 \, , \quad V_{\parallel} = - {1 \over 2}V_0 + {R
   \over 2}{dV_0 \over d R} \, .
\end{equation}
Eqs.~(\ref{57}), (\ref{A3}), (\ref{A4}) and (\ref{A2}) yield an analytic
parametrization of the velocity dependent potential $V_2$ in terms of
$\alpha_s$ and the string tension $\sigma$.
Combining the static
and velocity dependent contributions, eqs.~(\ref{48}) and (\ref{55}), to the
effective Lagrangian $L_I$, we have,
\begin{equation} \label{58}
   L_I = - V_0 (R) - V_2 \, .
\end{equation}
For $g =0$,  Eqs.
(\ref{49}), (\ref{56}), and (\ref{58}) give,
\begin{equation} \label{59}
   L_I = {4 \over 3}{\alpha_s \over R} - {2 \over
  3}{\alpha_s \over
     R} \left \{ \vec v_1 \cdot \vec v_2 + {\vec v_1 \cdot \vec R \vec v_2
	\cdot \vec R \over R^2} \right \} \equiv L_D \, ,
\end{equation}
where $L_D$ is the Darwin Lagrangian (multiplied by the
color factor ${ 4 \over 3 } = 2 \tr Y^2$) describing the interaction of
electrically charged particles to second order in their velocities \cite{8}.
For $g \neq 0$ $L_I$ has the following behavior:
\begin{eqnarray} \label{60}
    L_I \longrightarrow L_D\, ,&  & \quad \hbox{as} \, R \rightarrow 0\, ,\\
    & & \nonumber \\
   L_I \longrightarrow - \sigma R \left \{ 1 -{ [ \vec R \times ( \vec v_1  +
   \vec v_2 )]^2 \over 8 R^2} \right \} & + A {(\vec R \times { d \vec R \over
d
   t})^2 \over 4 R} \, , & \quad \hbox{as} \, R \rightarrow \infty \, ,
   \label{61}
\end{eqnarray}
where $A$ is a constant determined from the large $R$ behavior of $V_+$,
i.e.,
\begin{equation} \label{62}
V_{+} (R) \longrightarrow - AR \, , \quad \quad \quad \hbox{as} \, R
\rightarrow \infty \, .
\end{equation}
 From eq.~(\ref{A3}) we see that
\begin{equation}\label{A5}
   A \approx .208 \sigma\, .
\end{equation}
In obtaining eq.~(\ref{61}), we have made use of eqs.~(\ref{57}), (\ref{A2}),
(\ref{A3}),
and (\ref{A4}).
\section{\bf  DISCUSSION}

\noindent {\bf a)} The Lagrangian $ L_I $, eq.~(\ref{58}), describes the quark
anti-quark interaction at
distances $R$ greater than the flux tube radius, $R_{FT} \sim .4fm$. (See
Sections VI and V f).
At shorter distances this
interaction is described by a Lagrangian $ L_{PT} $ obtained from perturbation
theory.  The Lagrangian $ L_{PT}$ is essentially the Darwin Lagrangian $L_D$,
eq.~(\ref{59}),
with $\alpha_s$ replaced by the running coupling constant.
Since $ L_I
\rightarrow L_D $ as $ R \rightarrow 0 $ , one can obtain a Lagrangian
applicable at all distances by using eq.~(\ref{58}) for $R > R_{FT}$ and using
eq.~(\ref{59}) for $ R < R_{FT}$  with $\alpha_s$ replaced by the running
coupling
constant.  Therefore the elimination of the gauge fields involved in the
construction of $L_I$ provides a natural framework in which to connect the
short
distance perturbative domain described by $A_\mu^a$ to the long distance domain
described by the $C_\mu^a$.
However, in our preliminary fits of energy levels
\cite{2}, we have been using $L_I$ at all distances.  This should
overestimate the short distance contributions.  Future calculations using
$L_{PT}$ at short distances should somewhat modify the values of $\alpha_s$ and
$\sigma$ obtained by fitting data with $L_I$ alone.  Comparison of these two
calculations will give an indication of the relative importance of long and
short distance contributions to energy levels. \bigskip

\noindent {\bf b)} Bardinelli, Montaldi, and Prosperi \cite{9} (BMP) have given
formal expressions
for the QCD
velocity dependent potentials in terms of generalized Wilson loops extending
previous work of Eichten and Feinberg \cite{10} who had written down analogous
expressions for spin-spin and spin orbit potentials.  Starting with their
general formulae and making use of Lorentz invariance, BMP obtained two
relations between the various amplitudes appearing in the velocity dependent
potential.  We now show that relativistic invariance of the Lagrangian density
${\cal L}_{\hbox{\tiny eff}}$ yields eqs.~(\ref{A4}) determining the potentials
$V_{-}$ and $V_{\parallel}$ in terms of the
central potential $V_0$.  Eqs. (\ref{A4}) are equivalent to the relations
obtained by BMP using generalized
Wilson loops.  To derive eqs.~(\ref{A4}) set $\vec v_1 = \vec v_2 = \vec v$ in
eq.~(\ref{57})
so that the quark and anti-quark move with the same velocity.  Then ${ d \vec
R \over d t } = 0$, and eq.~(\ref{57}) for $V_2$ and hence eq.~(\ref{58}) for
$L_I$ reduces to
\begin{equation} \label{64}
   - L_I = V_0 (R) + V_{-} (R) v^2 + ( V_{\parallel} (R) - V_{-} (R)) {( \vec R
\cdot
   \vec v )^2 \over R^2} \, .
\end{equation}
Now consider an inertial frame $S^{'}$ where the quarks are at rest so that
${\cal L}_2^{'} = 0$.  Since ${\cal L}_{\hbox{\tiny eff}} = {\cal L}_0 + {\cal
L}_2$ is a
scalar, ${\cal L}_0 + {\cal L}_2 = {\cal L}^{'}_0$ .  Then from eq.~(\ref{41}),
\begin{equation} \label{65}
   - L_I = - \int d \vec x {\cal L}_0^{'} = - \int d \vec x^{'} \sqrt{1 - v^2}
{\cal
   L}_0^{'} = \sqrt{1 - v^2}  V_0 (R') \, ,
\end{equation}
where $R'$ is the distance between the quark anti-quark pair in $S'$ .  To
order $v^2$ , $R' = R + {1 \over 2} ({ \vec R \cdot \vec v \over R} )^2 R$ ,
and eq.~(\ref{65}) becomes
\begin{equation} \label{66}
   - L_I = ( 1 - {v^2 \over 2}) \left \{ V_0 (R) + {1 \over 2} ({ \vec R \cdot
\vec v
   \over R })^2 R { d V_0 \over d R } \right \} \,.
\end{equation}
Comparison of eqs.~(\ref{66}) and (\ref{64}) then give the relations
(\ref{A4}). \bigskip

\noindent {\bf c)} To see the meaning of the remaining terms in $V_2$
choose $\vec v_1 + \vec v_2 = 0$ in eq.~(\ref{57}).  Then
\begin{equation} \label{67}
   V_2 = {V_{+} (R) \over 4 R^2} (\vec R \times {d \vec R \over d t})^2 - {4
   \alpha_s \over 3 R} {( \vec R \cdot {d \vec R \over d t})^2 \over 4 R^2}
   \, .
\end{equation}
Note that the coefficient of the radial oscillation $ ( \vec R \cdot {d \vec
R \over d t})^2$ comes solely from the Biot Savart contribution (\ref{56}).
This has a simple physical
interpretation.  For a pure radial oscillation ${d \vec R \over d t} \propto
\vec R $ \, , i.e. $ \vec R \times { d \vec R \over d t} = 0$ , and only the
second(Biot Savart) term in eq.~(\ref{67}) survives.  This is to be expected
since
confining forces operate primarily perpendicular to the direction of the flux
tube and provide no hindrance to motion along the axis of the flux tube.

To isolate the $V_{+}$ contribution to eq.~(\ref{67}),
consider the special case of circular motion.
$ { d \vec R \over d t } = \vec \omega \times \vec R $, $\vec \omega \cdot \vec
R = 0
$.  Then eq.~(\ref{67}) becomes
\begin{equation} \label{69}
   V_2 = {1 \over 4} R^2 V_{+} (R) \omega^2 = - {1 \over 2} I (R) \omega^2 \, ,
\end{equation}
where
\begin{equation} \label{70}
   I(R) \equiv - {1 \over 2} R^2 V_{+} (R)
\end{equation}
   is the moment of inertia
of the rotating quark anti-quark pair. The constant $A$ (see eqs.~(\ref{62})
and (\ref{A5})), determines $I(R)$ for large $R$.  The minus sign in
eq.~(\ref{69})
arises because $V_2$ defined by eq.~(\ref{55}) is the mechanical potential
energy rather than the field energy $E_{FIELD}$.  We have calculated
$E_{FIELD}$ and also $\vec J_{FIELD}$, the
angular momentum of the color field distribution produced
by a quark anti-quark pair undergoing circular motion of frequency $\vec
\omega$.  As expected, we find:
\begin{equation} \label{71}
   E_{FIELD} = { 1 \over 2} I(R)\omega^2\, , \quad
   \vec J_{FIELD} = I (R) \vec \omega \, ,
\end{equation}
where $I (R)$ is given by eq.~(\ref{70}).  Since $ \vec J_{FIELD} $ includes
contributions from both scalar fields $ B_i$
and vector fields $ C_\mu$ , it automatically accounts for the spin of the
dual gluon and hence differs essentially from that calculated in
phenomenological
flux tubes models. \bigskip

\noindent  {\bf d)}
A Monte Carlo lattice gauge theory calculation of the QCD velocity dependent
potentials starting
from the formulae of BMP would provide a direct check of our predictions for
$ V_2$.  Such a calculation would complement
the Monte Carlo calculations of Huntley and Michael \cite{11} of the spin orbit
potentials which gave data consistent with the predictions
of ${\cal L}_{\hbox {\tiny eff}}$ augmented by spin dependent terms \cite{2}.
These spin dependent potentials like $V_2$ involved only $\alpha_s$ and
$\sigma$ as parameters.
\bigskip

\noindent  {\bf e)} Starting from $L_I$ augmented by the spin dependent terms
and non-relativistic quark kinetic energy terms, we have constructed the
Hamiltonian $H$ for a heavy quark anti-quark system by the canonical
procedure.  We then determined the parameters $\alpha_s$, $\sigma$, $m_c$ and
$m_b$ by finding a best fit to the 17 known levels of $c \bar c$ and $ b \bar
b$ systems. The best fit of the parameters were \cite{1}:
\begin{description}
\item[] (i) $\alpha_s = .3515$
\item[] (ii) $\sigma = .214 \,\hbox{GeV}^2$
\item[] (iii) $ m_c = 1.317\, \hbox{GeV}$
\item[] (iv) $ m_b = 4.747\, \hbox{GeV}$
\end{description}
Using $g^2/\lambda = 5$ and eq.~(\ref{A1}) gives the values $\lambda = 1.788$
and $B_0 = 94.4 \, \hbox{MeV}$ for the parameters appearing directly in
${\cal L}_{\hbox{\tiny eff}}$.  These parameters then give a value $M^2 \equiv
6g^2 B_0^2 = (680 \, \hbox{MeV})^2$ for the dual gluon mass squared, (see eq.~
(\ref{44})), and a value $M_B = 477 \hbox{MeV}$ for the lightest Higgs meson.
This latter mass determines the order of magnitude of the the flux tube
radius. (See the following section).
\bigskip

\noindent {\bf f)} Although the main purpose of this paper was the construction
of an effective
quark Lagrangian which contained no explicit reference to the underlying
effective local fields, as noted in (c) one can also use ${\cal L}_{\hbox{\tiny
eff}}$ to
calculate local quantities such as the energy momentum tensor $T_{\mu\nu}
(x)$.  We have calculated the energy density $ T_{00} (x)$ surrounding a
static quark anti-quark pair separated by a distance $R$.  The energy density
on the median plane $ z = 0 $ falls off exponentially with the distance $\rho$
from the flux tube axis, which is compatible with the results of the
calculation of analogous quantities in lattice gauge theory \cite{12}.
Although the
correspondence between local quantities calculated from
Monte Carlo data is not completely clear and the comparison is to some extent
qualitative, it provides a check on the physical picture underlying $L_I$.
Such tests are not even in principle possible for purely phenomenological
quark anti-quark potentials.

The Monte Carlo calculations \cite{12} were carried
out in $SU(2)$ gauge theory for an interquark spacing $ R \sim 1.5 f m $ and
gave a root mean squared flux tube radius $ R_{FT} \sim .2 fm$.  The energy
distribution calculated from ${\cal L}_{\hbox{\tiny eff}}$ broadens as a
function of $R$.
For $ R = .7 fm$.  $R_{FT} = .3 fm$, for $ R = 1.3 fm$, $R_{FT} = .4 fm$, and
for
$R = 2.6 fm$, $R_{FT} = .45 fm$, at which point the ``flux tube'' is close to
its
maximum width. \bigskip

\noindent {\bf g)} S. Maedan and T. Suzuki \cite{13} on the basis of somewhat
different reasoning have also
proposed a Lagrangian describing long distance QCD in terms of dual
potentials.  The classical equations of motion derived from their Lagrangian
do not differ substantially from the equations obtained from ${\cal L}_{\hbox
{\tiny eff}}$ in the gauge we have chosen.  Kamizawa, Matsubara, Shiba and
Suzuki \cite{14} have solved the static equations of their theory in the
presence of
three
static quarks and have calculated the static potential $V_0 (\vec x_1\, ,\vec
x_2\, ,\vec x_3)$ for three quarks in a baryon.
The essential complication of the baryon problem is that the source term
$\partial^\mu G^s_{\mu\nu}$ in eq.~(\ref{33}) involves a string connecting the
three quarks.
Starting with the static solution to eqs.~(\ref{33}) and (\ref{35}) one can
solve
these equations to first order in the velocity and insert the result into $
{\cal L}_2$, eq.~(\ref{40}).  Carrying out this procedure would yield a unique
velocity dependent
potential $ V_2 = - \int {\cal L}_2 d \vec x$ and an effective interaction
Lagrangian $ L_I ( \vec x_1 \,, \vec x_2 \,, \vec x_2 \,;\, \vec v_1 \, , \vec
v_2
\, , \vec v_3 )$ for three quarks inside a baryon.

\begin{center}
{\bf VI. REGGE TRAJECTORIES}
\end{center}

To  treat mesons containing light quarks and in particular to find Regge
trajectories for particles composed only of light quarks it is essential to
use relativistic kinematics.
The Lagrangian describing a quark anti-quark pair of masses $ m_1$ and $m_2$
respectively is
\begin{equation} \label{72}
   L = - m_1 \sqrt{1 - v_1^2} - m_2 \sqrt{1 - v_2^2} - V_0 -V_2 \, .
\end{equation}
The Lagrangian (\ref{72}) accounts for the kinematic effects of relativistic
velocities but cannot account for inelastic processes.  We saw that it was
possible to define a velocity dependent potential $V_2$ only to second order
in the velocity.
As in electrodynamics, the whole concept of a potential
dependent only upon particle positions and coordinates has meaning only to
order $v^2$ because of the appearance of radiation to order $(v^2)^2$.
We know for relativistic velocities $ v^2 \sim 1$ that it is essential to
use the relativistic form of the kinetic energy.  It makes sense to use
eq.~(\ref{72}) for relativistic velocities provided the energies are
sufficiently low so that inelastic processes are not important.
Radiation occurs only for frequencies $\omega$ which are greater than the
masses of the dual gluon or Higgs field.  These masses are of the order of
500 MeV. (See Sec. V e).  The inverse of these masses determine $R_{FT}$,
the radius of the flux tube.  Hence there is no radiation for frequencies $
\omega < 1 / R_{FT}$.  On the other hand for quarks separated by a
distance $R$, $\omega R < 1$.  Therefore at distances $R > R_{FT}$,
$ \omega < {1 \over
R_{FT}}$ and there is no radiation.  The dominant relativistic effects are then
kinematic and are included in the Lagrangian (\ref{72}).

To calculate the energy of states having large angular momentum, we can use
classical dynamics.  We find the leading Regge trajectory, i.e. the energy and
angular momentum of those states
which have the largest angular momentum for a given energy, by considering
circular orbits so that
\begin{displaymath}
 \vec v_1 \cdot \vec R = \vec v_2 \cdot \vec R = 0.
\end{displaymath}
Then the Lagrangian $L$, eqs.~(\ref{72}), (\ref{58}), and (\ref{57}) reduces to
\begin{equation} \label{73}
   L = - m_1 \sqrt{1 - v_1^2} - m_2 \sqrt{ 1 - v_2^2} - V_0 - {V_{+} \over 4}
   (\vec v_1 - \vec v_2)^2 - {V_{-} \over 4} ( \vec v_1 + \vec v_2 )^2 \,.
\end{equation}
 From eq.~(\ref{73}) we calculate the canonical momentum at $ \vec p_i =
{\partial L \over \partial \vec v_i}$ , $i = 1 \, ,2$ , the energy
$E = \vec p_1 \cdot \vec v_1 + \vec p _2 \cdot \vec v_2 - L$ and the angular
momentum $ \vec J = \vec x _1 \times \vec p _1 + \vec x_2 \times \vec p_2$. We
find,
\begin{equation} \label{74}
   E = {m_1 \over \sqrt{1 - v_1^2}} + {m_2 \over \sqrt{ 1 - v_2^2}} + V_0 -
   V_+ {( \vec v_1 - \vec v_2)^2 \over 4} - V_{-} {( \vec v_1 + \vec v_2 )^2
   \over 4} \, ,
\end{equation}
and
\begin{equation} \label{75}
   \vec J = {\vec R \over 2} \times \left[ {m_1 \vec v_1 \over \sqrt{1 -
   v_1^2}} - {m_2\vec v_2 \over \sqrt{1 - v_2^2}} - V_{+} ( \vec v_1 - \vec
v_2)
   \right] \, .
\end{equation}
In obtaining eq.~(\ref{75}) we have made use of the fact that we are working in
the
center of mass frame $ \vec p_1 = - \vec p_2$.

Using Hamilton's principle,
the Lagrangian (\ref{73}), and the fact that we have circular orbits so that
$ \vec v_1 \cdot \dot{\vec v}_1 = \vec v_2 \cdot \dot{\vec v}_2 = 0 $ , we
obtain the equation of motion
\begin{eqnarray} \label{76}
  {m_1 \dot{\vec v}_1 \over \sqrt{1 - v_1^2}} - {(V_{+} + V_{-}) \over 2}
  \dot{\vec v_1} - {(V_{-} - V_{+}) \over 2} \dot{\vec v_2}
 & =   \nonumber \\
 & & \nonumber \\
    - {\vec R \over R} \left[ {\partial V_0 \over \partial R} + {(\vec v_1 -
   \vec v_2)^2 \over 4} { \partial V_{+} \over \partial R} +
   {(\vec v_1 + \vec v_2)^2 \over 4} { \partial V_{-} \over \partial R} \right]
   &  \, ,
\end{eqnarray}
along with a corresponding equation obtained by replacing $m_1$ by $m_2$ and
interchanging $\vec v_1$ and $\vec v_2$ in eq.~(\ref{76}). For circular motion
of frequency $\vec{\omega}$, $ \vec v_1
= \vec{\omega} \times \vec R_1$ and $ \vec v_2 = \vec{\omega} \times \vec R_2$
where $ \vec R_1 = R_1 \hat{R}$, $\vec R_2 = - R_2 \hat{R}$, and $ R_1 + R_2 =
R$.  We can then write eq.~(\ref{76}) in the form
\begin{eqnarray} \label{77}
   \omega^2 \left[ {m_1 R_1 \over \sqrt{ 1 - \omega^2 R_1^2}} - {(V_{+} +
   V_{-}) R_1 \over 2} + {(V_{-} - V_{+}) R_2 \over 2} \right] & = &
   \nonumber\\
   & & \nonumber \\
    {\partial V_0 \over \partial R} + {\partial V_{+} \over \partial R}
   ({\omega^2 R^2 \over 4}) + \omega^2 {\partial V_{-} \over \partial R }{(R_1
-R_2)^2
   \over 4} \, , & &
\end{eqnarray}
along with the corresponding equation obtained from eq.~(\ref{77}) by
replacing $m_1$ by $m_2$ and interchanging $R_1$ and $R_2$.  Eqs. (\ref{77})
determine $R_1$ and $R_2$ in terms of $\omega$.  Substituting these
expressions for $R_1$ and $R_2$ into eqs.~(\ref{74}) and (\ref{75}) give
$\vec J$ and $E$ as functions of $\omega$, which upon elimination of $R$ yield
the leading Regge trajectory.

For the equal mass case, $m_1 = m_2 = m$, $R_1 =R_2 = {R \over 2}$, and
eqs.~(\ref{77}), (\ref{74}) and (\ref{75}) reduce to
\begin{equation} \label{78}
   {m \over \sqrt{1 - {\omega^2 R^2 \over 4}}}  = {2 \over \omega^2 R^2}
   R {\partial V_0 \over \partial R } + V_{+} + {\partial V_{+} \over \partial
   R} { R \over 2 } \, ,
\end{equation}
\begin{equation} \label{79}
   E = { 2 m \over \sqrt{ 1 - {\omega ^2 R^2 \over 4 }}} + V_0 - V_{+} { R^2
   \omega^2 \over 4 } \, ,
\end{equation}
\begin{equation} \label{80}
   \vec J = {R^2 \vec{\omega} \over 2 } \left[ { m \over \sqrt{ 1 - {\omega^2
   R^2 \over 4}}} - V_{+} \right] \, .
\end{equation}
The Regge trajectory $J$ as a function of $E^2$ obtained from these equations
is plotted in Fig. (3) for the cases of $m = 100 \,\hbox{MeV}$ and $m = 300
\,
\hbox{MeV}$.

We note that for a constituent quark mass of $100 \, \hbox{MeV}$,
this trajectory passes through the two known Regge recurrences of the $\pi$
meson.  For any quark mass the trajectory curves so that there are no
particles with $E^2$ negative.
For large $E^2$
the trajectory becomes linear with a slope $\alpha^{'}$ which is determined by
the behavior of $V_0$ and $V_{+}$ at large $R$, eqs.~(\ref{50}) and (\ref{62}).
For large $R$, $ { \omega^2 R^2 \over 4 } \sim 1$ and eq.~(\ref{78}) becomes
\begin{eqnarray} \label{81}
   { m \over \sqrt{ 1 - {\omega^2 R^2 \over 4 }}} & = & { 1 \over 2} \sigma R -
   A R - { A R \over 2} = { 1 \over 2 } ( \sigma - 3A ) R \, . \\
     R & \rightarrow & \infty  \nonumber
\end{eqnarray}
Note that the presence of the velocity dependent potential $V_{+}$ tends to
destabilize the system, and there are solutions to eq.~(\ref{81}) for $\omega$
only for $A < \sigma / 3$.  This condition is satisfied by the result
(\ref{A5}) of our numerical calculation.

Inserting the equilibrium condition eq.~(\ref{81}) into eqs.~(\ref{79}) and
(\ref{80}) yields the following expressions for $E$ and $\vec J$ valid at
large $R$
\begin{equation} \label{82}
   E = 2 R ( \sigma - A )\, , \quad  \vec J = \hat{\omega} {R^2 \over 2 } (
\sigma
   - A ) \, .
\end{equation}
Eq. (\ref{82}) then yields a linear Regge trajectory $ | \vec J | = \alpha^{'}
E^2$ with
\begin{equation} \label{83}
   \alpha^{'} = { 1 \over 8 \sigma ( 1 - {A \over \sigma})} \approx { 1 \over
   \sigma ( 6.34)}\, .
\end{equation}
Note the relativistic Salpeter Equation (with a central potential $V_0$) gives,
for large $l$,
linear Regge trajectories with $ \alpha^{'} = { 1 \over 8 \sigma}$.
This agrees with our classical result for $V_{+} = 0$.  Including the
velocity dependent potential then increases the slope of the Regge
trajectories.  Using the value $ \sigma = .214 \hbox{GeV}^2$, (see Sec. V e),
we obtain
\begin{equation} \label{84}
 \alpha^{'} = { 1 \over 1.36 }\, \hbox{GeV}^{-2} = .74
 \, \hbox{GeV}^{-1} \, .
\end{equation}

In collaboration with Lewis Fulcher we are now in the process of calculating
the energy levels of mesons containing
light quarks using a Hamiltonian obtained from the Lagrangian (\ref{71}).  The
velocity dependent terms in the Hamiltonian are uniquely determined from the
canonical procedure applied to $V_2$ (\ref{57}).  Since these velocity
dependent terms will be more important in mesons containing light quarks, the
comparison with experiment will be more sensitive to our prediction for $V_2$.
Furthermore, since the relevant distance scales will be larger than in mesons
containing only heavy quarks, the long distance part of the potential will
give a corresponding larger contribution to the energy levels.  For both these
reasons we feel that these calculations of energy levels of mesons containing
light quarks will provide a sensitive test of the theory.  The only new
parameters are the constituent masses of the light quarks.
\begin{center}
\newpage
{\bf VI. SUMMARY AND COMMENTS}
\end{center}

The original quark model hypothesized the existence of constituent quarks.  It
left the nature of their interaction an open question.  We have argued that
dual potentials are the natural variables describing long distance QCD and have
written down an effective Lagrangian ${\cal L}_{\hbox{\tiny eff}}$ determining
their interaction with quarks.  Since dual potentials describe only long
distance QCD they cannot have a local interaction with the fundamental quarks
of QCD.  We therefore assume that the dual potentials in ${\cal L}_{\hbox{\tiny
eff}}$ interact with constituent quarks.  The electromagnetic interaction can
be used to define electrons and positrons as the source of photons.  Similarly
the interaction mediated by dual potentials can be used to define constituent
quarks as the source of dual gluons.  Solving the field equations for these
dual potentials in the presence of a quark anti-quark pair leads
to an effective Lagrangian for constituent quarks inside mesons,
containing $\alpha_s$ and $\sigma$ as parameters.  Physically
this Lagrangian is the non abelian analogue of the effective Lagrangian
describing the interaction of a monopole anti-monopole pair in a
superconductor.  It thus provides a concrete realization of the Mandelstam
'tHooft dual superconducting mechanism for confinement which can now be
confronted by experiment.

\newpage
\begin{center}
{\bf FIGURE CAPTIONS}
\end{center}
\begin{description}

\item{Fig. 1} Dirac string connecting oppositely charged particles.\\
\item{Fig. 2} Diagram representing string cancellation mechanism of
eq.~(\ref{13}).\\
\item{Fig. 3}  The points represent the particles on the $\pi\, ,
\rho\, , \omega\, ,  and\, \phi$ Regge trajectories, and the solid lines are
the corresponding linear fits.\\
$\quad\quad$Dotted lines are Regge trajectories calculated from
eqs.~(\ref{78}),
(\ref{79}) and (\ref{80}): \\
$\quad\quad- \, - \, - \quad m_q = 100 \, \hbox{MeV} \, ,\quad \quad
\cdot\,\cdot\,\cdot\quad m_q =
300 \, \hbox{Mev}$.
\end{description}

\end{document}